\documentclass[12pt]{article}
\usepackage{amsfonts}
\usepackage{amsmath}
\usepackage{amssymb}
\usepackage{graphicx}
\usepackage{color}
\usepackage[left=1cm,top=1cm,bottom=1cm,right=1cm,nohead,nofoot]{geometry}

\def \be {\begin{equation}}
\def \ee {\end{equation}}
\def \bea {\begin{eqnarray}}
\def \eea {\end{eqnarray}}
\def \nn {\nonumber}

\def \rr {\raise.35ex\hbox{\small $\prime$}\kern-.17em{\mbox{\large $\imath$}}}
\def \del {\partial}
\def \dels {\partial\kern-.6em /\kern.1em}
\def \As {{A\kern-.5em / \kern.5em}}
\def \Ds {D\kern-.7em / \kern.5em}
\def \a {\alpha}

\def \b {\beta}

\def \g {\gamma}

\def \d {\delta}
\def \eps {\epsilon}

\def \ks {k\kern-.5em /}
\def \ls {l\kern-.5em /}

\def \II {I\hspace{-.1em}I\hspace{.1em}}

\def \IIA {\mbox{\II A\hspace{.2em}}}

\def \dd {\dot{\delta}}

\def \dm {\dot{\mu}}
\def \dn {\dot{\nu}}

\def \dlam {\dot{\lambda}}
\def \ds {\dot{\sigma}}
\def \dr {\dot{\rho}}

\reversemarginpar
\begin{document}
\begin{titlepage}

\begin{center}

\hfill
\vskip .2in

\textbf{\LARGE
Effective Action for D$p$-Brane \\
in Large RR $(p-1)$-Form Background \\
\vskip.5cm
}

\vskip .5in
{\large
Pei-Ming Ho$\,{}^{a,b,c,}$\footnote{e-mail address: pmho@phys.ntu.edu.tw} and
Chen-Te Ma$^{a,}$\footnote{e-mail address: yefgst@gmail.com}\\
\vskip 3mm
}
{\sl
${}^a$
Department of Physics and Center for Theoretical Sciences, \\
${}^b$
Center for Advanced Study in Theoretical Sciences, \\
${}^c$
National Center for Theoretical Sciences, \\
National Taiwan University, Taipei 10617, Taiwan,
R.O.C.}\\
\vskip 3mm
\vspace{60pt}
\end{center}
\begin{abstract}

We construct the low energy effective action for the bosonic sector
on a D$p$-brane in large constant RR $(p-1)$-form field background. 
The action is invariant under both $U(1)$ gauge symmetry 
and the volume-preserving diffeomorphism characterizing the RR-field background.
Scalar fields representing transverse coordinates of the D$p$-brane are included.
It also respects T-duality and is consistent with the action 
for M5-brane in $C$-field background.

\end{abstract}

\end{titlepage}

\section{Introduction}
\label{1}

It is well known that the classical low energy effective theory of D-branes 
can be described by the DBI action in curved spacetime \cite{Leigh},
and the low energy effective quantum theory of D-branes in flat spacetime 
is described by the super Yang-Mills action \cite{Witten:1995im}.
Interactions of D-brane with RR fields are described 
by the Wess-Zumino terms ${\cal L}_{WZ} = e^{F}C$.

In the low energy limit,
the dynamics on the D-brane is dominated by 
the zero modes of open strings ending on the D-brane,
although the exact theory includes higher order terms 
corresponding to the interactions mediated by higher oscillation modes of the open strings,
as well as those mediated by other D-branes.
When there is a large longitudinal NS-NS $B$-field background, 
the effective theory of a D-brane is deformed to a non-commutative gauge theory
\cite{Chu-Ho,Seiberg:1999vs},
as a result of the fact that the $B$-field background changes 
the zero modes of open strings,
so that higher derivative terms are turned on.
Similarly, when there is a large longitudinal $C$-field background in M theory,
the effective theory of M5-brane is a Nambu-Poisson gauge theory
\cite{M51,M52,Ho:2009zt}.
The interaction mediated by open membranes under the effect of the $C$-field 
is characterized by the gauge symmetry of volume-preserving diffeomorphism, 
where the volume-form is defined by the $C$-field.
Through the duality between M theory and type \IIA superstring, 
one can derive the D4-brane theory in large RR 3-form field background 
from M5-brane theory in large $C$-field background 
\cite{Ho,Ho:2012dn,Ma:2012dn}.
From the viewpoint of the D4-brane,
the volume-preserving diffeomorphism characterizes 
the interaction mediated by D2-branes in the RR 3-form $C$ field background.

In general, we expect that the large RR $(p-1)$-form field 
turns on certain contributions of D$(p-2)$-branes to 
the interactions on the D$p$-brane worldvolume theory, 
which is characterized by a volume-preserving diffeomorphism 
for a volume $(p-1)$-form defined by the RR field background.
In this work,
we focus on the bosonic part of the Lagrangina, 
and extend the results about D4-brane \cite{Ho} in large RR 3-form field 
to D$p$-brane in large RR $(p-1)$-form field.
Conceptually simple, this is in practice a nontrivial task 
due to the complexity of the highly nonlinear structure of gauge symmetry. 
In the end of the construction, 
the bosonic effective action for a D$p$-brane in large longitudinal RR $(p-1)$-form background 
in flat spacetime has the following characteristic features.

\begin{enumerate}
\item
The Lorentz symmetry of the 10-dimensional spacetime 
in the presence of the D$p$-brane is broken by the RR field
to the subgroup $SO(1,1)\times SO(p-1)\times SO(9-p)$.
\item
The D$p$-brane dynamics is characterized by 
the gauge symmetry of the $(p-1)$-form-volume-preserving diffeomorphism
with a $(p-2)$-form gauge potential.
\item
The $U(1)$ gauge symmetry of the D$p$-brane is still present.
Part of the $U(1)$ gauge field is electric-magnetic dual to the gauge field for 
volume-preserving diffeomorphism.
\item
The D$p$-brane action is related to D$(p\pm 1)$-brane action via T-duality.
\item
At the leading order in the large RR field background, 
the effective action formally agrees with that for the trivial background
(but with a different metric).
This is analogous to the situation for constant $B$-field background.
(The noncommutativity vanishes when $B$ is infinite.)
\end{enumerate}

We construct the effective action by requiring the first 4 properties, 
starting with the D4-brane theory given by \cite{Ho,Ma:2012dn}.
The most informative constraint comes from the requirement of T-duality.
The last property is carried over from D4-brane in $B$-field background
through dualities.
The final form of the bosonic action is presented in eqs.(\ref{L})--(\ref{L2}).
We leave the fermionic part of the action and supersymmetry for future investigations.

Let us now specify the range of parameters for which 
the effective action obtained in the work gives a good approximation.
First we recall that 
the limit for the Nambu-Poisson structure to dictate 
world-volume interactions on the M5-brane in $C$-field background 
is given as \cite{Furuuchi}
\bea
&\ell_p \sim \eps^{1/3}, \\
&g_{\mu\nu} \sim 1, \quad (\mu, \nu = 0, 1, 2) \\
&g_{\dm\dn} \sim \eps, \quad (\dm, \dn = 3, 4, 5) \\
&C_{\dm\dn\dlam} \sim \eps^0
\eea
with $\eps \rightarrow 0$.
This is consistent with the double scaling limit of Seiberg-Witten 
\cite{Seiberg:1999vs} for the non-commutative gauge theory 
to be a good low energy effective theory for a D-brane in 
large NS-NS $B$-field background, 
when we compactify, say, the $x^5$-direction.
The {\em open-membrane} metric for the M5-brane in 
the large $C$-field limit is approximated by
\be
G_{\dm\dn} = \frac{1}{8}(2\pi)^4 \ell_p^6 
g^{\dm_1\dn_1}g^{\dm_2\dn_2}C_{\dm\dm_1\dm_2}C_{\dn\dn_1\dn_2},
\ee
which remains finite in the limit $\eps \rightarrow 0$.
If we compactify the $x^2$-direction, 
instead of the $x^5$-direction,
one can derive the scaling limit for D4-brane in RR $C$-field background 
to be characterized by the Nambu-Poisson bracket \cite{Ho}
via M theory/\IIA superstring duality.
Then we can see through T-dualities that the scaling limit 
for the effective theory given in this paper 
to be a good approximation for a D$p$-brane in 
RR $(p-1)$-form background is given by 
\bea
&\ell_s \sim \eps^{1/2},
\qquad
g_s \sim \eps^{-1/2}, 
\\
&g_{\a\b} \sim 1, \quad (\a, \b = 0, 1)
\\
&g_{\dm\dn} \sim \eps, \quad (\dm, \dn = 2, 3, \cdots, p)
\\
&C_{\dm_1\cdots\dm_{p-1}} \sim 1
\eea
with $\eps \rightarrow 0$.

\section{Gauge Symmetry}
\label{gaugesymmetry}

In the presence of the RR $(p-1)$-form background,
the D$p$-brane world-volume is naturally decomposed into 
the product of 2-dimensional Minkowski space
and a $(p-1)$-dimensional Euclidean space 
with the global symmetry $SO(1,1)\times SO(p-1)$.
Correspondingly, 
we will use $x^{\a}$ ($\a = 0, 1$) and $y^{\dm}$ ($\dm =2, 3, \cdots, p$)
to denote world-volume coordinates of the D$p$-brane.
The RR $(p-1)$-form background 
\be
C^{(p-1)} = \frac{1}{(p-1)!} C_{\dm_1\cdots\dm_{p-1}} dy^{\dm_1} \cdots dy^{\dm_{p-1}}
= \frac{1}{g} dy^2 \cdots dy^p
\ee
defines a VPD (volume-preserving dffeomorphism)
generated by a $(p-1)$-bracket
\be
\{f_1, f_2, \cdots, f_{p-1}\} \equiv
\eps^{\dm_1 \dm_2 \cdots \dm_{p-1}} 
(\del_{\dm_1}f_1) (\del_{\dm_2}f_2) \cdots (\del_{\dm_{p-1}}f_{p-1}).
\ee
(The value of $p$ can be $2, 3, \cdots, 9$)
The 1-bracket is just an ordinary derivative, 
generating translation (as the length-preserving diffeomorphism).
The 2-bracket is a Poisson bracket
generating area-preserving diffeomorphism.
The 3-bracket is the simplest generalization of Poisson bracket
and it is often called the Nambu-Poisson bracket.
In general, the $(p-1)$-bracket satisfies the {\em generalized Jacobi identity}
\bea
\{f_1, \cdots, f_{p-2}, \{g_1, \cdots, g_{p-1}\}\}
&=&
\{\{f_1, \cdots, f_{p-2}, g_1\}, \cdots, g_{p-1}\}\}
+ 
\nn \\
&+&
\{g_1, \{f_1, \cdots, f_{p-2}, g_2\}, \cdots, g_{p-1}\}\}
+\cdots
\nn\\
\cdots&+&
\{g_1, \cdots, g_{p-2}, \{f_1, \cdots, f_{p-2}, g_{p-1}\}\},
\label{GJI}
\eea
as a generalization of the Jacobi identity for the Poisson bracket.

The gauge symmetry for a D$p$-brane in large RR $(p-1)$-form background 
was proposed in Ref.\cite{Ho} to be $U(1) \times$ VPD.
We will refer to a field $\Phi$ as {\em VPD-covariant} if it transforms under VPD as
\be
\d\Phi = \{f_1, f_2, \cdots, f_{p-2}, \Phi\} = \kappa^{\dm}\del_{\dm}\Phi,
\ee
where 
\be
\kappa^{\dm} = 
\eps^{\dm_1\cdots\dm_{p-2}\dm}(\del_{\dm_1}f_1)\cdots(\del_{\dm_{p-2}}f_{p-2})
\ee
is the VPD parameter and it is divergenceless
\be
\partial_{\dot\mu}\kappa^{\dot\mu}=0.
\ee
Due to the generalized Jacobi identity (\ref{GJI}), 
the $(p-1)$-bracket of $(p-1)$ VPD-covariant fields $\Phi_i$,
\be
\{\Phi_1, \cdots, \Phi_{p-1}\},
\ee
is also VPD-covariant.

The $(p-2)$-form gauge potential $b_{\dm_1\cdots\dm_{p-2}}$ for VPD
is more conveniently described as a vector field in the $(p-1)$-dimensional subspace
\be
b^{\dm_1} = \frac{1}{(p-2)!}\eps^{\dm_1 \dm_2 \cdots \dm_{p-1}} 
b_{\dm_2\cdots\dm_{p-1}}.
\ee
The potential $b$ is not VPD-covariant,
and its transformation law under VPD can be 
derived by demanding that 
\be
X^{\dm} \equiv \frac{y^{\dm}}{g} + b^{\dm}
\ee
be VPD-covariant \cite{Ho}.
The VPD-covariant field strength ${\cal H}$ can be defined as \cite{Ho}
\be
{\cal H}^{\dm_1\dm_2\cdots\dm_{p-1}} \equiv 
g^{p-2}\{X^{\dm_1}, X^{\dm_2}, \cdots, X^{\dm_{p-1}}\}-\frac{1}{g}\epsilon^{\dm_1\dm_2\cdots\dm_{p-1}}.
\ee
It has a single independent component ${\cal H}_{23\cdots p}$.

Now we turn to the more familiar $U(1)$ gauge symmetry on a D-brane.
The $U(1)$ field strengths on a D$p$-brane are modified 
in order for them to be VPD-covariant.
Denoting the $U(1)$ potential as $a_A$ $(A = 0, 1, 2, \cdots, p)$,
we define the $U(1)$ field strengths by 
\bea
{\cal F}_{\dm\dn}&\equiv&
\frac{g^{p-3}}{(p-3)!} \eps_{\dm\dn\dm_1\cdots\dm_{p-3}}
\{X^{\dm_1}, \cdots, X^{\dm_{p-3}}, a_{\dot{\rho}}, y^{\dot{\rho}} \},
\label{Fdmdn1} \\
{\cal F}_{\a\dm}&\equiv&
{V^{-1}}_{\dm}^{~\dn}(F_{\a\dn}+gF_{\dn\dd}\hat{B}_{\a}^{~\dd}), 
\label{Fadm1} \\
{\cal F}_{\a\b}&\equiv&
F_{\a\b}+g[-F_{\a\dm}\hat{B}_{\b}^{~\dm}-
F_{\dm\b}\hat{B}_{\a}^{~\dm}+
gF_{\dm\dn}\hat{B}_{\a}^{~\dm}\hat{B}_{\b}^{~\dn}],
\label{Fab1}
\eea
where $F_{AB} \equiv \del_A a_B - \del_B a_A$ is usual Abelian field strength,
and $V_{\dm}{}^{\dn}$ and $\hat{B}_{\a}{}^{\dm}$ are defined by
\bea
V_{\dn}^{~\dm}&\equiv&\d_{\dn}^{~\dm}+g\del_{\dn}b^{\dm}, \\
M_{\dm\dn}{}^{\a\b} &\equiv&
V_{\dm\dr}V_{\dn}{}^{\dr}\d^{\a\b}-g\eps^{\a\b}F_{\dm\dn},
\\
\hat{B}_{\a}{}^{\dm} &\equiv&
(M^{-1})^{\dm\dn}{}_{\a\b}
(V_{\dn}^{~\ds}\del^{\b}b_{\ds}+\eps^{\b\g}F_{\g\dn}
+g\del_{\dn}X^I {\cal D}^{\b}X^I).
\label{def-B}
\eea
It is straightforward to check that all field strengths ${\cal F}_{AB}$ are 
VPD-covariant and invariant under $U(1)$ gauge transformations.
The last term in (\ref{def-B}) was absent in Ref.\cite{Ho} 
because the scalar fields $X^I$ were omitted. 
This term was computed for the D4-brane in Ref.\cite{Ma:2012dn},
and here we generalize it to D$p$-brane by demanding T-duality.

Another difference from the notation in earlier works
Refs.\cite{Ho,Ma:2012dn}
is that there the field strength ${\cal F}_{\dm\dn}$ was defined by
\bea
{\cal G}_{\dm\dn} &\equiv& 
F_{\dm\dn}+g[\del_{\ds}b^{\ds}F_{\dm\dn}-
\del_{\dm}b^{\ds}F_{\ds \dn}-\del_{\dn}b^{\ds}F_{\dm \ds}]
\nn \\
&=&V^{~\dr}_{\dr}F_{\dm\dn}+V^{~\dr}_{\dm}F_{\dn\dr}+
V^{~\dr}_{\dn}F_{\dr\dm},
\label{Gdmdn}
\eea
instead of (\ref{Fdmdn1}). 

The quantity ${\cal G}_{\dm\dn}$ 
(denoted as ${\cal F}_{\dm\dn}$ in Refs.\cite{Ho,Ma:2012dn})
is identical to ${\cal F}_{\dm\dn}$ for $p \leq 4$, 
but they are different for $p > 4$ at higher orders in $g$.
Roughly speaking, 
the ambiguity in the choice of a covariant field strength is due to 
the presence of the additional gauge potential $b^{\dm}$.
It turns out that ${\cal G}_{\dm\dn}$ is not as convenient as ${\cal F}_{\dm\dn}$
for the sake of T-duality considerations.

For the convenience of the reader,
we list here the gauge transformation laws 
with the transformation parameter $\Lambda = (\lambda, \kappa)$ 
for the fields in a D$p$-brane theory \cite{Ho}:
\bea
\delta_{\Lambda}X^I
&=&g \kappa^{\dm}\partial_{\dm}X^I,
\\
\delta_{\Lambda} b^{\dm}&=&
\kappa^{\dm}+g\kappa^{\dn}\del_{\dn}b^{\dm}\label{transf-bdm},\\
\delta_{\Lambda} a_{A}&=&
\del_{A}\lambda+
g(\kappa^{\dn}\del_{\dn}a_{A}+a_{\dn}\del_{A}\kappa^{\dn}),
\label{transf-adm}\\
\delta_{\Lambda} \hat{B}_{\a}^{~\dm} &=& 
\del_{\a}\kappa^{\dm}+
g(\kappa^{\dn}\del_{\dn}\hat{B}_{\a}^{~\dm}-
\hat{B}_{\a}^{~\dn}\del_{\dn}\kappa^{\dm})
\label{transf-B}.
\eea
Here we listed the transformation law for the composite field $\hat{B}_{\a}{}^{\dm}$ together 
with those for the fundamental fields $X^I, b^{\dm}, a_A$ 
because of its importance of playing the role of a gauge potential.
Together with $b^{\dm}$, 
they allow us to define VPD-covariant derivatives,
\bea
{\cal D}_{\alpha}X^I&\equiv&
\partial_{\alpha}X^I-g\hat{B}_{\alpha}{}^{\dot\mu}\partial_{\dot\mu}X^I, \\
{\cal D}_{\dot\mu_1}X^I&\equiv&
\frac{(-1)^p}{(p-2)!}g^{p-2}\epsilon_{\dot\mu_1\dot\mu_2 \cdots \dot\mu_{p-1}}
\{X^{\dot\mu_2}, X^{\dot\mu_3}, \cdots, X^{\dot\mu_{p-1}}, X^I\}.
\eea

In addition to the covariant derivatives and field strengths,
a class of VPD-covariant and $U(1)$-invariant quantities is given by 
\bea
{\cal O}_{nml} \equiv
\{X^{\dm_1}, \cdots, X^{\dm_n}, 
a_{\dn_1}, \cdots, a_{\dn_m}, \frac{y^{\dn_1}}{g}, \cdots, \frac{y^{\dn_m}}{g}, X^{I_1}, \cdots, X^{I_l}\},
\label{Onml} 
\\
\mbox{where} \quad n, m, l \geq 0 \quad \mbox{and} \quad n + 2m + l = p-1.
\nn
\eea
In fact, 
${\cal D}_{\dm}X^I$, ${\cal H}_{2\cdots p}$ and ${\cal F}_{\dm\dn}$
all belong to this class.
Since both $X^{\dm}$ and $X^I$ are VPD-covariant, 
it is obvious that (\ref{Onml}) is VPD-covariant when $m = 0$.
It is interesting that the combination $(a_{\dm}, y^{\dm})$ is also covariant 
when it appears in the $(p-1)$-bracket as a pair.
As the coordinate $y^{\dm}$ is not a dynamical variable, 
it cannot transform under VPD.
But if it were covariant, 
it should transform like 
$\d y^{\dm} = g\kappa^{\dn}\del_{\dn}y^{\dm} = g\kappa^{\dm}$.
On the other hand,
the VPD-transformation of $a_{\dm}$
has just the precise additional term
(the third term in (\ref{transf-adm}))
in addition to the covariant piece (the second term in (\ref{transf-adm}))
to compensate the non-covariance of $y^{\dm}$.
The relevant identity is
\be
(\del_{[\dm}\d_{\Lambda} a_{\dr})(\del_{\dn]} y^{\dr}) 
= g(\del_{[\dm}(\kappa\cdot\del a_{\dr}))(\del_{\dn]}y^{\dr})
+ g(\del_{[\dm}a_{\dr})(\del_{\dn]}\kappa^{\dr}).
\ee
The left hand side is the part of a $(p-1)$-bracket 
relevant to the gauge transformation of the pair $(\d_{\Lambda} a_{\dm}, y^{\dm})$, 
and the right hand side is what we would have for this part of the $(p-1)$-bracket
if both $a_{\dm}$ and $y^{\dm}$ were VPD-covariant.

\section{Lagrangian from T-duality}

The requirement of T-duality imposes a strong constraint on 
the Lagrangian for the D$p$-brane.
In this section 
we use the T-duality as the major guideline to build 
the effective Lagrangian for a D$p$-brane in RR field background.

Upon compactification on a circle in the direction of $y^{\dm=p}$,
the D$p$-brane is dual to a D$(p-1)$-brane in the T-dual theory.
The gauge potential $a_{\dm=p}$ is T-dual to a scalar field $X^{I=p}$
interpreted as a transverse coordinate for the D$(p-1)$-brane. 
\footnote{
There is an ambiguity in the symbol $X^p$. 
It could mean $X^{\dm=p} = \frac{y^p}{g} + b^p$ on the D$p$-brane 
or $X^{I=p}$ as a scalar field representing a transverse coordinate of the D$(p-1)$-brane 
which is T-dual to the D$p$-brane.
Hence we will use the notation $X^{\dm=p}$ versus $X^{I=p}$ to avoid ambiguity.
}
The gauge potential $b^{\dm=p}$ can be set to zero as a choice of gauge fixing.
(Both the number of components of $b^{\dm}$ and that of VPD gauge transformations 
reduce by one via T-duality.)
The D$(p-1)$-brane action can be derived from the D$p$-brane action 
by the replacement
\bea
a_p &\rightarrow& X^{I = p}, 
\label{apXp} \\
b^p &\rightarrow& 0, 
\label{bp0} \\
\del_p &\rightarrow& 0 \quad \mbox{when acting on the fields}.
\label{dp0}
\eea

Since the definition of field strengths and covariant derivatives depend 
on the dimension of the D-brane,
we use a superscript in parenthesis,
$(p)$ or $(p-1)$ on quantities defined 
for a D$p$-brane or a D$(p-1)$-brane, respectively,
to avoid confusion.
Then one can check that (\ref{apXp})--(\ref{dp0}) imply 
\be
\hat{B}^{(p)}_{\a}{}^{\dm} \rightarrow \hat{B}^{(p-1)}_{\a}{}^{\dm}, 
\qquad
\hat{B}^{(p)}_{\a}{}^p \rightarrow \eps_{\a}{}^{\b} {\cal D}^{(p-1)}_{\b} X^{I=p},
\ee
and then we have the following rules for T-duality transformation
\bea
{\cal D}^{(p)}_{\a} X^I &\rightarrow& {\cal D}^{(p-1)}_{\a} X^I, 
\label{Da-T} \\
{\cal D}^{(p)}_{\dm} X^I &\rightarrow& {\cal D}^{(p-1)}_{\dm} X^I, \\
{\cal D}^{(p)}_p X^I &\rightarrow& 0, \\
{\cal F}^{(p)}_{\dm\dn} &\rightarrow& {\cal F}^{(p-1)}_{\dm\dn}, \\
{\cal F}^{(p)}_{\dm p} &\rightarrow& {\cal D}^{(p-1)}_{\dm} X^{I=p}, \\
{\cal F}^{(p)}_{\a \dm} &\rightarrow& {\cal F}^{(p-1)}_{\a \dm}, \\
{\cal F}^{(p)}_{\a p} &\rightarrow& {\cal D}^{(p-1)}_{\a} X^{I=p}, \\
\frac{1}{2}\eps^{\a\b}{\cal F}^{(p)}_{\a\b} &\rightarrow& 
\frac{1}{2}\eps^{\a\b}{\cal F}^{(p-1)}_{\a\b} - g({\cal D}^{(p-1)}_{\a}X^{I=p})^2, \\
{\cal H}^{(p)}_{23\cdots p} &\rightarrow& {\cal H}^{(p-1)}_{23\cdots (p-1)},
\label{H-T}
\eea
where $\dm, \dn \neq p$.

Starting with the D4-brane case \cite{Ho} 
which is derived from the M5-brane theory in $C$-field background
\cite{M51,M52,Ho:2009zt},
we can straightforwardly go down to D3-brane and D2-brane theories via 
the T-duality transformation rules listed above.
To go up to D5-brane, 
we need to look for Lagrangians that would reduce to the D4-brane theory 
through the replacements (\ref{Da-T})--(\ref{H-T}).
This is more complicated but doable. 
Similarly we can climb all the way up to D9-brane.
It is a nontrivial consistency check that 
the global symmetry $SO(1,1)\times SO(p-1)\times SO(9-p)$ 
is respected for all D$p$-branes.
It is also not obvious whether it will be possible to express the Lagrangian such that 
it takes a compact, universal form for all D$p$-branes.

Instead of showing the details of taking T-dualities 
and the trial and error to rewrite the Lagrangian in a compact, 
manifestly covariant form,
we just give the final expression of the Lagrangian 
and show that it does respect T-duality as required.
The effective Lagrangian of a D$p$-brane
in large RR $(p-1)$-form background is
\be
{\cal L} = {\cal L}_1 + {\cal L}_2,
\label{L}
\ee
where
\be
{\cal L}_1 \equiv 
- \frac{1}{2}({\cal D}_{\a}X^I)^2 + \frac{1}{2g}\eps^{\a\b}{\cal F}_{\a\b}
+ \frac{1}{2}{\cal F}_{\a\dm}^2,
\label{L1}
\ee
and 
\bea
{\cal L}_2 &\equiv&
- \frac{1}{2} \sum_{n,m,l \in S}
\frac{g^{2(p-2-m)}}{(n!)(m!)^2(l!)} 
\{ X^{\dm_1}, \cdots, X^{\dm_n}, a_{\dn_1}, \cdots, a_{\dn_m}, y^{\dn_1}, \cdots, y^{\dn_m}, 
X^{I_1}, \cdots, X^{I_{l}} \}^2
\nn \\
&=& - \frac{g^{2(p-2)}}{2} \sum_{n,m,l \in S} 
C^{p-1}_{nmm}{\cal O}_{nml}^2,
\label{L2}
\eea
where ${\cal O}_{nml}$ is defined in (\ref{Onml}),
the indices $(n, m, l)$ are to be summed over the set
\be
S \equiv \{(n, m, l) \; |\; n, m, l \geq 0 \; ; \; n+2m+l = p-1\},
\ee
and the coefficient is defined by
\footnote{
Note that over the set $S$, $C^{p-1}_{nmm} = C^{p-1}_{nml}$.
}
\be
C^{q}_{nml} \equiv \frac{1}{n!m!l!(q-n-m-l)!}.
\ee
In short, up to an overall factor,
${\cal L}_2$ is a sum over all the VPD-covariant quantities in the class (\ref{Onml}) squared,
naturally weighed by a combinatorial factor.
The VPD-covariant quantities that do not belong to the class (\ref{Onml}) 
are collected in ${\cal L}_1$ in a peculiar way on which we will have more comments.

Apparently,
each term in ${\cal L}_1$ and ${\cal L}_2$ are $U(1)$-invariant and VPD-covariant.
Furthermore, ${\cal L}_1$ and ${\cal L}_2$ are invariant under T-duality by themselves.
Upon using the T-duality transformation rules in (\ref{Da-T})--(\ref{H-T}),
we have 
\bea
{\cal L}^{(p)}_1 &=& 
- \sum_{I=p+1}^9 \frac{1}{2}({\cal D}^{(p)}_{\a}X^I)^2 
+ \frac{1}{2g}\eps^{\a\b}{\cal F}^{(p)}_{\a\b}
+ \sum_{\dm=2}^p \frac{1}{2}{\cal F}^{(p)2}_{\a\dm}
\nn \\
&\rightarrow&
- \sum_{I=p+1}^9 \frac{1}{2}({\cal D}^{(p-1)}_{\a}X^I)^2 
+ \left[\frac{1}{2g}\eps^{\a\b}{\cal F}^{(p-1)}_{\a\b} - ({\cal D}^{(p-1)}_{\a}X^p)^2\right]
\nn \\
&&
+ \left[\sum_{\dm=2}^{p-1} \frac{1}{2}{\cal F}^{(p-1)2}_{\a\dm}
+ \frac{1}{2} ({\cal D}^{(p-1)}_{\a}X^p)^2\right]
= {\cal L}^{(p-1)}_1.
\eea
There is a cancellation between the T-duality transformations
of the second and third terms of ${\cal L}^{(p)}_1$,
such that the kinetic term $({\cal D}^{(p-1)}_{\a}X^{I=p})^2$ 
has the right coefficient to join the kinetic term of $X^I$ in ${\cal L}^{(p-1)}_1$.

Regarding the T-duality transformation of ${\cal L}_2$,
the key observation is that 
the fields $X^{\dm}, X^I, a_{\dm}$ $(\dm \neq p)$
living on the D$(p-1)$-brane are independent of $y^p$,
so the $(p-1)$-bracket does not vanish only if 
one of the slots are taken by $y^{\dm = p}$
(or $X^{\dm = p}$).
Hence we have
\bea
{\cal L}^{(p)}_2 
&=& 
- \frac{g^{2(p-2)}}{2} \sum_{n,m,l}
C^{p-1}_{nmm}
\{ X^{\dm_1}, \cdots, X^{\dm_n}, a_{\nu_1}, \cdots, a_{\nu_m}, 
\frac{y^{\nu_1}}{g}, \cdots, \frac{y^{\nu_m}}{g}, 
X^{I_1}, \cdots, X^{I_{l}} \}_{(p-1)}^2 
\nn \\
&\rightarrow&
- \frac{g^{2(p-2)}}{2} \sum_{n,m,l}
C^{p-1}_{nmm} \times 
\nn \\
&& \times
\Big[
n
\{ X^{\dm_1}, \cdots, X^{\dm_{n-1}}, X^{\dm=p}, a_{\nu_1}, \cdots, a_{\nu_m}, 
\frac{y^{\nu_1}}{g}, \cdots, \frac{y^{\nu_m}}{g}, 
X^{I_1}, \cdots, X^{I_{l}} \}_{(p-1)}^2
\nn \\
&&
+ m^2 
\{ X^{\dm_1}, \cdots, X^{\dm_n}, a_{\nu_1} \cdots, a_{\nu_{m-1}}, X^{I=p}, 
\frac{y^{\nu_1}}{g}, \cdots, \frac{y^{\nu_{m-1}}}{g}, \frac{y^p}{g}, 
X^{I_1}, \cdots, X^{I_{l}} \}_{(p-1)}^2
\Big]
\nn \\
&=&
- \frac{g^{2(p-3)}}{2} \sum_{n,m,l} \Big[
\frac{1}{(n-1)!(m!)^2 l!} 
\{ X^{\dm_1}, \cdots, X^{\dm_{n-1}}, a_{\nu_1}, \cdots, a_{\nu_m}, y^{\nu_1}, \cdots, y^{\nu_m}, 
X^{I_1}, \cdots, X^{I_{l}} \}_{(p-2)}^2
\nn \\
&&
+
\frac{1}{n!((m-1)!)^2 l!} 
\{ X^{\dm_1}, \cdots, X^{\dm_n}, a_{\nu_1} \cdots, a_{\nu_{m-1}}, y^{\nu_1}, \cdots, y^{\nu_{m-1}},  
X^{I_1}, \cdots, X^{I_{l}}, X^{I=p} \}_{(p-2)}^2
\Big]
\nn \\
&=&
{\cal L}^{(p-1)}_2
\eea
after imposing the T-duality transformations (\ref{Da-T})--(\ref{H-T}).
In the above, $I_i = p+1, \cdots, 9$, 
and we used $\{\cdots\}_{(p-1)}$ and $\{\cdots\}_{(p-2)}$
to denote the $(p-1)$-bracket and $(p-2)$-bracket 
for D$p$-brane and D$(p-1)$-brane, respectively.
For the last step, we relabeled $n$ as $(n+1)$ for the first term,
and we relabeled $m$ as $(m+1)$, and $l$ as $(l-1)$ for the second term.
This concludes the proof that ${\cal L}_2$ also respects T-duality.

The content of the Lagrangian (\ref{L}) needs some explanation.
The first term in ${\cal L}_1$ is the standard kinetic term for the scalar fields $X^I$,
apart from those terms $({\cal D}_{\dm}X^I)^2$ hidden in ${\cal L}_2$ (see (\ref{DX2})).
The second term in ${\cal L}_1$ is of the form of the Wess-Zumino term 
for the coupling between the field strength ${\cal F}_{01}$
and the RR field background $C_{2\cdots p} = 1/g$.
The third term in ${\cal L}_1$ looks like part of the standard kinetic term 
for the $U(1)$ gauge field ${\cal F}_{AB}$, 
but the other two terms $\frac{1}{4}{\cal F}_{\a\b}^2$ and $\frac{1}{4}{\cal F}_{\dm\dn}^2$
of the kinetic terms are missing.
While the latter is hidden in ${\cal L}_2$ (see (\ref{F2})), 
the former is hidden in the Wess-Zumino term and ${\cal L}_2$ in a nontrival way
that we will explain below.
In fact, 
the coefficient of the third term in ${\cal L}_1$ has the wrong sign that 
will be corrected by the contribution of the Wess-Zumino term.
The kinetic term $\frac{1}{2}{\cal H}_{2\cdots p}^2$ for the VPD gauge potential
is again hidden in ${\cal L}_2$ (see (\ref{H2})).

Some of the terms in ${\cal L}_2$ can be conveniently expressed 
in terms of covariant derivatives and field strengths.
The term corresponding to $m=0, n=p-1$ is
\be
{\cal L}_2^{(m=0, n=p-1)} = - \frac{1}{2(p-1)!}
({\cal H}_{\dm_1\cdots\dm_{p-1}} + \frac{1}{g}\eps_{\dm_1\cdots\dm_{p-1}})^2.
\label{H2}
\ee
Like the Wess-Zumino term in ${\cal L}_1$,
this expression also suggests that the background RR field has the magnitude $1/g$.

We also have 
\be
{\cal L}_2^{(m=0, n=p-2)} = - \frac{1}{2} ({\cal D}_{\dm}X^I)^2, 
\label{DX2}
\ee
and
\be
{\cal L}_2^{(m=1, n=p-3)} = - \frac{1}{4} {\cal F}_{\dm\dn}^2.
\label{F2}
\ee
These are the kinetic terms missing in ${\cal L}_1$.

To see how the kinetic term $\frac{1}{4}{\cal F}_{\a\b}^2$ is hidden 
in the Wess-Zumino term and ${\cal L}_2$,
we examine the perturbative expansion of the Lagrangian in powers of $g$.
To the 0-th order, 
$\hat{B}_{\a}{}^{\dm} \simeq \del_{\a}b^{\dm} + \eps_{\a\b}F^{\b\dm}$
and the Wess-Zumino term is
\bea
\frac{1}{2g}\eps^{\a\b}{\cal F}_{\a\b} 
&\simeq& 
\frac{1}{g}F_{01} - \eps^{\a\b}F_{\a\dm}\hat{B}_{\b}{}^{\dm} 
\nn \\ 
&\simeq&
\eps^{\a\b}(\del_{\b}F_{\a\dm})b^{\dm} - F_{\a\dm}^2 + \cdots
\nn \\
&\simeq&
- F_{01}H_{23\cdots p} - F_{\a\dm}^2 + \cdots,
\label{WZ-g}
\eea
where we ignored total derivatives 
and $H_{23\cdots p} = \del_{\dm}b^{\dm}$ is the 0-th order part of ${\cal H}_{23\cdots p}$.
Note that the second term in the last line flips the sign 
of the third term in ${\cal L}_1$ at the 0-th order.
To the 0-th order in $g$, 
the pure gauge field terms in the Lagrangian are 
\be
{\cal L}_{gauge} =
-\frac{1}{2}(H_{23\cdots p}+F_{01})^2
-\frac{1}{4}F_{AB}F^{AB}
+ \mbox{total derivatives} + {\cal O}(g).
\label{Sgauge10Dp}
\ee
The term $H_{23\cdots p}^2$ appears in ${\cal L}_2$ (\ref{H2}), 
and the cross term $F_{01}H_{23\cdots p}$ is found in the Wess-Zumino term in (\ref{WZ-g}).
Completing the square to get the first term in ${\cal L}_{gauge}$,
we compensate it by $\frac{1}{2}F_{01}^2$ that was missing.
Incidentally,
since $H_{23\cdots p}$ is the only component of the field strength 
for the potential $b^{\dm}$, 
and it has no time derivative terms,
we can integrate it out in the perturbation theory
and the Lagrangian reduces to 
that of Maxwell theory in $(p+1)$ dimensions.
In fact,
one may check that
the complete Lagrangian reduces to that of the ordinary D$p$-brane 
in trivial background.


\section{Conclusion and Outlook}
\label{4}

In this work we have constructed the bosonic part 
of the low energy effective action (\ref{L})--(\ref{L2}) 
for a single D$p$-brane in the large RR $(p-1)$-form potential background. 
The Lagrangians possess 
the interesting structure of a $(p-1)$-bracket characterizing the VPD gauge symmetry.
The existence of a universal expression of the Lagrangian consistent with T-duality 
is highly nontrivial,
and can be taken as a supporting evidence for the correctness of the result.
On the other hand, 
in our derivations we have ignored all total derivative terms.
The possibility of additional topological terms remains open.

The Lagrangian (\ref{L})--(\ref{L2}) is applicable to D$p$-brane in 
RR $(p-1)$-form background only for $p = 2, 3, \cdots, 9$
because the $(p-1)$-bracket is defined only for $p \geq 2$.
The RR 0-form (axion) background does not introduce interactions 
to the effective theory of a D$1$-brane as 
higher RR fields to higher dimensional D-branes, 
as it is suggested by the absence of a ``$0$-bracket''.

The case of D3-brane is of special interest because
a D3-brane in RR 2-form background is S-dual to 
a D3-brane in NS-NS $B$-field background, 
and the latter can be described as a noncommutative gauge theory.
In the large $B$-field background, 
the noncommutative structure can be approximated by the Poisson bracket.
In fact, the low energy effective theory of M5-brane in large $C$-field background 
has already been shown \cite{M52,Ho} to agree with 
both D4-brane in NS-NS $B$-field background 
and D4-brane in RR $C$-field background, 
depending on whether the direction of compactification is $x^{\dm=5}$ or $x^2$.
Since the D4-brane theories for the two different backgrounds reduce to
D3-branes in NS-NS $B$-field or RR field background
through dimensional reduction on $x^2$ or $x^{\dm=5}$, respectively,
it is more or less obvious that,
to the leading order (Poisson approximation),
the noncommutative $U(1)$ gauge theory is dual (equivalent) to 
the effective theory defined by (\ref{L}) for $p = 3$, 
as they are both dimensional reduction of the same M5-brane theory 
on a torus in the directions of $x^2$ and $x^{\dm = 5}$.
On the other hand, 
since the higher order terms beyond the Poisson approximation is known 
for the D3-brane in $B$-field background, 
it implies that we can also obtain corresponding higher order terms 
for the D3-brane in RR 2-form field background 
via the duality transformation.
This task is left for future works.

It is well known that the usual description of $B$ field via the Wess-Zumino terms
is related to the formulation using noncommutative gauge theory
via the Seiberg-Witten map \cite{Seiberg:1999vs}.
The analogous Seiberg-Witten map for the M5-brane in $C$-field background 
was constructed to all orders in Ref.\cite{Furuuchi}.
It should also be possible to write down the Seiberg-Witten map 
for D$p$-branes in RR $(p-1)$-form background.

A natural question is whether one can deform the $(p-1)$-bracket 
in a way analogous to how the Poisson bracket (2-bracket) 
is deformed by the Moyal bracket.
The higher order terms in the deformation is expected 
to capture higher order corrections 
in inverse powers of the background RR field.
However, there are no-go theorems \cite{Chen:2010ny}
saying that such generalization does not exist for 3-brackets.
The same argument can be easily generalized to higher brackets.
A proper generalization for the 3-bracket or higher brackets will 
probably call for an enhancement of the symmetry group, 
from the $(p-1)$-form-volume-preserving diffeomorphism 
to a larger group.
The 2-brackets for D3-brane can be deformed.
Information about higher order terms can be obtained 
through S-duality as commented above.

In this work we have not yet fully explored all consequences of T-duality.
Only the compactification in a longitudinal direction of the RR field ($y^{\dm}$)
is considered in this work.
It will be interesting to explore effective action for D-branes 
with different RR field backgrounds via T-duality in other directions.

Perhaps the most interesting problem is to understand 
the physical origin of the interaction characterized by the $(p-1)$-bracket 
as contributions of the quantum fluctuation of 
D$(p-2)$-branes ending on D$p$-branes.
Let us recall that
for D-branes in NS-NS $B$-field background, 
the noncommutative geometric nature of D-brane 
can be demonstrated in two different ways. 
One way is to quantize an open string ending on D-brane 
in the $B$-field background \cite{Chu-Ho}.
The other way is to examine correlation functions of 
open string vertex operators in the $B$-field background \cite{Schomerus:1999ug}.
For the RR field background,
the first approach was carried out in Ref.\cite{Chu:2010eb},
where a generalization of canonical formulation 
was motivated by manifest diffeomorphism invariance.
Nambu-Poisson bracket (3-bracket) is used in place of Poisson bracket 
in a generalized Hamiltonian formulation.
A generalization to higher brackets is straightforward.
The second approach was applied to study the correlation functions 
of open membranes in $C$-field background in M theory \cite{Ho:2007vk}.
This approach also leads to the appearance of Nambu-Poisson bracket.
A more careful analysis of either or both approaches 
might lead us to further understanding of higher order corrections 
to the Nambu-Poisson bracket.
We leave this interesting questions for future study.

\section*{Acknowledgement}

We thank Heng-Yu Chen, Wei-Ming Chen, Chong-Sun Chu, 
Takeo Inami, Pei-Wen Peggy Kao, Fech Scen Khoo, Yutaka Matsuo,
Hiroaki Nakajima and Chi-Hsien Yeh for discussions.
This work is supported in part by 
NTU (grant \#NTU-CDP-102R7708), 
and by National Science Council, Taiwan, R.O.C.

\end{document}